\documentclass[10pt, conference, compsocconf]{IEEEtran}
\IEEEoverridecommandlockouts
\usepackage{cite}
\usepackage{amsmath,amssymb,amsfonts}
\usepackage{algorithmic}
\usepackage{graphicx}
\usepackage{textcomp}
\usepackage{xcolor}
\usepackage{soul}
\usepackage{url}
\usepackage{fancyhdr, lipsum}
\usepackage{setspace}
\let\labelindent\relax
\usepackage{enumitem}
\usepackage[symbol]{footmisc}
\def\BibTeX{{\rm B\kern-.05em{\sc i\kern-.025em b}\kern-.08em
    T\kern-.1667em\lower.7ex\hbox{E}\kern-.125emX}}
\begin{document}

\title{HaoCL: Harnessing Large-scale Heterogeneous Processors Made Easy\vspace{-18pt}}

\author{
\IEEEauthorblockN{
Yao Chen\IEEEauthorrefmark{1},
Xin Long\IEEEauthorrefmark{2},
Jiong He\IEEEauthorrefmark{3},
Yuhang Chen\IEEEauthorrefmark{1},
Hongshi Tan\IEEEauthorrefmark{4}, \\
Zhenxiang Zhang\IEEEauthorrefmark{2},
Marianne Winslett\IEEEauthorrefmark{1},\IEEEauthorrefmark{5},
Deming Chen\IEEEauthorrefmark{1},\IEEEauthorrefmark{5}
}
\IEEEauthorblockA{
\IEEEauthorrefmark{1}Advanced Digital Sciences Center,
\IEEEauthorrefmark{2}Alibaba Group,
\IEEEauthorrefmark{3}Institute of High Performance Computing, A*Star\\
}
\IEEEauthorblockA{
\IEEEauthorrefmark{4}National University of Singapore,
\IEEEauthorrefmark{5}University of Illinois at Urbana-Champaign
}
\vspace{-20pt}
\thanks{*Yao Chen and Jiong He both made equal contributions to this work.}
}
\maketitle

\begin{abstract}
The pervasive adoption of Deep Learning (DL) and Graph Processing (GP) makes it a de facto requirement
to build large-scale clusters of heterogeneous accelerators including GPUs and FPGAs.
The OpenCL programming framework can be used on the individual nodes of such clusters but is not intended for deployment in a distributed manner.
Fortunately, the original OpenCL semantics naturally fit into the programming environment of  heterogeneous clusters.
In this paper, we propose a \textit{h}eterogeneity-\textit{a}ware \textit{O}pen\textit{CL}-like (\textit{HaoCL}) programming framework to facilitate the programming of a wide range of scientific applications including DL and GP workloads on large-scale heterogeneous clusters. With HaoCL, existing applications can be directly deployed on heterogeneous clusters without any modifications to the original OpenCL source code and without awareness of the underlying hardware topologies and configurations. 
Our experiments show that HaoCL imposes a negligible overhead in a distributed environment, and provides near-liner speedups on standard benchmarks when computation or data size exceeds the capacity of a single node.
The system design and the evaluations are presented in this demo paper.
\end{abstract}


\section{Introduction} \label{sec:introduction}
\begin{figure*}[ht]
    \centering
    \includegraphics[width=0.8\textwidth]{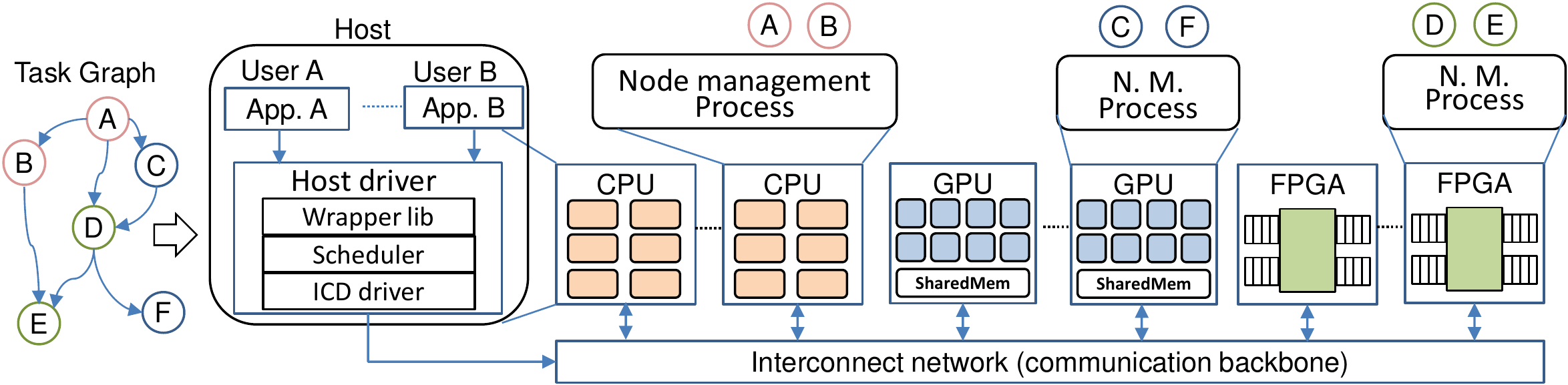}
    \vspace{-2mm}
    \caption{Framework overview.}
    \vspace{-6mm}
    \label{fig:system}
\end{figure*}

With the increasing volume of data to be processed and higher complexity of workloads in modern academic and industrial applications, large scale heterogeneous computing clusters have become an attractive solution that offer both massively parallel computing capability and large storage capacity by integrating traditional multi-core CPUs, many-core GPUs and power-efficient FPGAs into High Performance Computing (HPC) systems~\cite{haas, clouddnn, shufflefly}.
\bstctlcite{IEEEexample:BSTcontrol}
Though these hardware accelerators were originally designed for specific tasks, general-purpose programming languages have been released by their vendors so that complex applications can be mapped to such hardware by offloading tasks of applications partially or entirely. Languages such as CUDA and OpenCL are powerful tools for programmers to deploy applications on these accelerators. CUDA is a proprietary toolchain released by NVIDIA exclusively for NVIDIA GPUs. 
In contrast, OpenCL and OpenACC can be used in a cross-platform manner so that one copy of source code can run on different OpenCL-compatible devices~\cite{shufflefly, snucl2012, snucl_2016, vcl, rcuda}. 
However, most existing applications still target a single system equipped with one type of accelerator~\cite{rcuda, vcl, snucl2012, snucl_2016, shufflefly}.
We lack of solutions that can provide an efficient and easy-to-use abstraction for a heterogeneous cluster.

 
Modern scientific applications usually consist of a large number of sub-tasks that have different hardware resource preferences in order to optimize either performance, power consumption, or monetary cost.
However, power consumption and cost make it impractical to equip each node in a cluster with all types of accelerators.
Ad-hoc assignment of these sub-tasks to nodes without holistic deployment information about all accelerators across the cluster inevitably lead to serious under-utilization~\cite{vcl, rcuda, snucl2012, snucl_2016}. This problem becomes even worse in large-scale cloud systems that need to serve massive requests from many users simultaneously. 
Though some existing solutions achieve a unified abstraction of a heterogeneous cluster \cite{vcl, rcuda, snucl2012, snucl_2016}, 
they simply provide a wrapper of the native OpenCL implementation which offloads the tasks to remote devices via networking components, without trying to optimize resource allocations. 
Therefore, they cannot serve modern applications efficiently due to the stringent requirements on performance, power efficiency and usability.
\begin{itemize}[noitemsep, leftmargin=*, topsep=0pt]
    \item \textbf{Performance.} The heterogeneity of modern clusters requires the scheduler in the system to be aware of the device differences. Without a detailed device model, a scheduler for a heterogeneous cluster will produce sub-optimal scheduling plans and result in worse performance. 

    \item \textbf{Power efficiency.}
    CPU, GPU and FPGA resources in large-scale clusters provide different efficiency characteristics for different tasks. 
    For maximum power efficiency, a scheduler requires device model and run-time information of the tasks on the different devices.

    \item \textbf{Usability.}
    Although existing solutions targeting different computing devices can provide performance benefits, integrating them together efficiently still requires great effort. 
\end{itemize}

No existing approach provides both heterogeneity-aware and performance-oriented scheduling in an automated or user-guided manner.
Instead, they usually adopt a static and heterogeneity-oblivious mapping between tasks (OpenCL kernels) and hardware resources.
Such optimistic allocation patterns would result in serious resource under-utilization in scenarios where the number of users and accelerators is large in commercial clouds. 
Besides, FPGAs are moving into cloud computing as well~\cite{clouddnn, haas}. However, previously proposed frameworks \cite{snucl2012, snucl_2016, vcl, rcuda} only consider CPUs and GPUs. There are few works on how FPGAs can be utilized to improve the energy efficiency and how a heterogeneity-aware scheduling algorithm \cite{haas} should be designed with additional hardware types.

To solve the above questions and meet the stringent requirements of modern complex and compute-intensive applications, we propose a
heterogeneity-aware framework (HaoCL) 
that can efficiently abstract and manage different types of resources including FPGAs in heterogeneous HPC clusters in a holistic manner and also reduce the burden on application designers.
In addition, designers can design and illustrate their own scheduling algorithms and embed them into HaoCL to achieve their performance objectives.
In summary, HaoCL makes the following contributions.

\begin{itemize}[noitemsep, leftmargin=*, topsep=0pt]
    \item 
    A framework that manages different types of accelerators in a holistic manner with fine-grained resource scheduling capability.
    \item 
    An extensible run-time resource monitoring and scheduling component that supports both built-in and user customized scheduling policies.
    \item 
    A light-weight and easy-to-extend communication backbone built on an optimized asynchronous communication middleware which is specifically designed for a large-scale distributed environment.
    \item 
    Support for the same application programming interfaces (APIs) as OpenCL (or OpenCL-like for FPGAs), which significantly reduces the integration and migration overhead of current applications.
\end{itemize}
\section{Related Works} \label{sec:related}

VCL~\cite{vcl} allows applications on one hosting-node to transparently utilize cluster-wide devices (CPUs and/or GPUs); however, the coarse-grained granularity from workload limits the flexibility of it. RCUDA~\cite{rcuda} provides a unified abstraction of the clusters equipped with GPUs. However, it only supports the CUDA API as well as NVIDIA GPUs. It fails to provide a unified view of other types of accelerators such as AMD CPU/GPU and FPGAs. 
Specifically, the abstraction of FPGAs is even more challenging due to its special computing paradigms. High Level Synthesis (HLS) with the CUDA and OpenCL as input provides an efficient approach to implement and optimize applications on FPGAs~\cite{fcuda, shufflefly, xpilot, lopass}, but it still targets single device.

SnuCL~\cite{snucl2012} and SnuCL-D~\cite{snucl_2016} support OpenCL to achieve cross-platform function. However, their lack of multi-user support and very coarse-grained scheduling among workloads and resources prohibit the full utilization of the devices. 
HaaS~\cite{haas} is a heterogeneous-aware streaming processing framework that optimizes multiple runs on tasks to different devices in the system. The topology level scheduling and the heavy streaming back-end have limited its wide adoption in modern complex applications.

Complex applications consist of different tasks with different computation patterns, which suit different computational accelerators. Hence, a finer-granularity description of the tasks and flexible scheduling of them onto different devices (processor/accelerator) is critical for fully utilizing the resources in a large-scale heterogeneous cluster.
Although OpenCL provides fine-grained support of task description, state-of-the-art frameworks do not provide flexible scheduling, 
and only support limited hardware types, which motivate our HaoCL work in this study.

\section{System Design} \label{sec:system}

Our framework wraps native OpenCL APIs so that existing applications can be deployed on a cluster of heterogeneous resources naturally with few modifications. As such, we follow the abstract execution model of OpenCL but enhance it with heterogeneity-aware optimizations. 

\subsection{Framework Overview} \label{subsec:overview}
The overall architecture design of the HaoCL framework is shown in Fig. \ref{fig:system}. 
It consists of a single host node and multiple device nodes (CPU, GPU or FPGA). Each device node contains its own CPU, and the CPU for each of the GPU and FPGA device node is omitted for easy understanding. The nodes are connected through Gigabit Ethernet. The host node executes the host program in an OpenCL application and is also responsible for the message packaging and message delivering across the entire cluster. Note that the FPGA is used as a streaming processor with different performance characteristics from CPU or GPU. HaoCL framework is simplified to consist of only three major functional components: 1) an \textbf{OpenCL wrapper library and task scheduling component}, 
2) a \textbf{communication backbone} and 3) a \textbf{node management process} on each child node. The task scheduling component is aware of the hardware architecture of the individual nodes and delivers the kernel tasks with consideration of the run-time information of the kernel on the nodes.

\subsection{OpenCL-compatible driver}
HaoCL implements a full set of standard OpenCL API calls to make it compatible with existing applications implemented in OpenCL. Specifically, it consists of an OpenCL wrapper lib and an extendable scheduling component.

\paragraph{OpenCL Wrapper Lib} 
The OpenCL Wrapper Lib adopts identical names as standard OpenCL APIs to maintain good usability and portability. For each OpenCL API call, the same API in HaoCL creates a message package that contains the information of the function's name and arguments. Besides, the wrapper lib also creates data packages containing all data in OpenCL buffers that have been called in this API and sends it to the specified compute node through the communication backbone. The device nodes unpack the message and execute it with their own environment.
To support different hardware platforms with their own drivers, Installable Client Driver (ICD) is provided as the common entry point to those specific drivers on device nodes after intercepting those calls in applications implemented with the OpenCL Wrapper Lib. We extend the original ICD to be compatible with the front-end wrapper layer and the communication backbone for remote API calls forwarding so that each call to the standard OpenCL APIs can be executed in the form of implementation according to the remote devices and vendor drivers.

\paragraph{Extendable task scheduling component} 
The task scheduling component on the host node is aware of the run-time information of the accelerator nodes. In the current version, it delivers the kernel tasks to device nodes based on users' instructions. However, it is designed in an extendable manner so that it can be upgraded to an automatic scheduler with the runtime profiling information from the cluster to enable more accurate heterogeneity-aware task scheduling.

\subsection{Communication Backbone} \label{subsec:comm}
The communication backbone is built as the network among host and devices for data transfers. The design of it imposes great challenges due to the requirement of flexibility and efficiency as well as the support of complex data transfer patterns.
To address these challenges, we leverage the Boost library to take advantages of its strengths~\cite{boost}.
1) Its stability and scalability ensure the correctness of the transmission of high volume data.
2) Its flexibility of extension enables adding new nodes easily.
3) The asynchronous input and output feature enables an efficient unblocking data transfer style which is necessary for a distributed cluster. 
4) Moreover, it can be more easily utilized to implement complex inter-node data transfer schemes in the OpenCL API.

When the Node Management Process (NMP) in the device node is created, it obtains the {IP} address and port from the user's input, then uses Boost's \textit{acceptor} structure as a message and data listener.
Once the listener is created, it listens to the port asynchronously. When messages/data comes, it creates a thread to read and unpacking the incoming message, then starts listening to the port again.

The communication method of the host process is almost the same as the node process (it also has the message and data listeners), but it reads the address and port defined in a system configuration file and creates a message and a data listener for each node. 
Besides, the listener will listen to the port synchronously. So after the host process sends a message through the message listener, it will wait for the response message and then take the next action according to the type of the message.

The communication backbone takes care of the message delivering among multiple different accelerator nodes. We leverage a mapping mechanism to address this. When the user program calls the \textit{clGetDeviceIDs} function, the wrapper lib will create a device {ID} request message for each compute node. After the response message from the node process is received, the backbone obtains the device's id of each compute node and records this mapping. 

\subsection{Node Manage Process} 
The daemon process runs on each device (accelerator) node for the actual execution of OpenCL API calls. It receives the commands from the workload scheduler along with additional information such as user ID, device ID, shared flag (whether a device is shared or not) and the number of resources from those explicit requests, and parses them for compilation and execution. Especially, the FPGA nodes do not have flexible support for task allocation. So the tasks are pre-built as executable binaries with the bitstreams to provide customized hardware solution together with bandwidth optimization~\cite{shufflefly}.
\section{Experiments} \label{sec:exp}


\begin{table}
\centering
\vspace{-2mm}
\caption{Benchmark Applications.}
\label{tab:benchmark}
\vspace{-2mm}
\resizebox{0.5\textwidth}{!}{%
\begin{tabular}{|l|l|l|}
\hline
App. & Description & In. size \\ \hline
MatrixMul & Matrix multiplication  & 760MB \\ \hline
CFD &Unstructured grid finite volume solver  & 800MB  \\ \hline
kNN &Finds k-nearest neighbors in unstructured data set & 100MB \\ \hline
BFS &Traverses all the connected components in a graph & 240MB \\ \hline
SpMV & Sparse matrix-vector multiplication in CSR format &1.1GB \\ \hline
\end{tabular}%
}
\vspace{-6mm}
\end{table}

We select a set of commonly used benchmarks for Machine Learning (ML) and Graph Processing (GP) for a comprehensive evaluation of our design.
The high demand for computations and data storage in the ML and GP applications requires high computing and storage capability that cannot be provided by a single computing node. HaoCL is designed to leverage the great computing capability of the large-scale distributed heterogeneous accelerators to process such applications.

\subsection{Evaluation Setup} \label{subsec:expsetup}


\emph{Environmental setup.}
We conduct our evaluations in the Alibaba Cloud with the elastic computing nodes. 
All nodes are equipped with Intel Xeon E5-2686 CPU. Each GPU node has an NVIDIA Tesla P4 GPU and each FPGA node has a Xilinx VU9P FPGA. 16 GPU nodes and 4 FPGA nodes are involved in our evaluations.
All the nodes are connected through Ethernet.
Specifically, for the heterogeneity evaluation, we construct clusters with different scales consists with different numbers of different types of devices.


\emph{Benchmark setup.} We select the representative OpenCL benchmarks from different sources: Rodinia~\cite{rodinia} and SHOC~\cite{shoc}.
The characteristics of the applications and their input sets are summarized in Table~\ref{tab:benchmark}.

\subsection{End-to-end Improvement} \label{subsec:end2endresults}

Fig.~\ref{fig:e2e_speedup} shows the scalable performance of different applications towards a single GPU/FPGA or GPU+FPGA node with a native OpenCL environment. 
We schedule the tasks with different numbers and types of nodes, denoted as GPU, FPGA and GPU + FPGA.
Overall, HaoCL could achieve a linear performance improvement compared to the single node implementation of the benchmarks when the number of nodes scales. However, the performance improvement also depends on the computation pattern and communication characteristics of the benchmarks.
We also compare our HaoCL framework to the Snu{CL}-D framework~\cite{snucl_2016}.

Our framework shows better performance improvement when the number of nodes scales for a wide range of benchmarks.
In addition, HaoCL supports the scheduling of clusters built with hybrid GPU and FPGA nodes and still achieves linear performance improvement when the node number increases. Note CFD cannot be implemented on SnuCL-D without significant change.

\subsection{Heterogeneity Evaluation} \label{subsec:heterogeneity}
\begin{figure}
    \centering
    \includegraphics[width=\columnwidth]{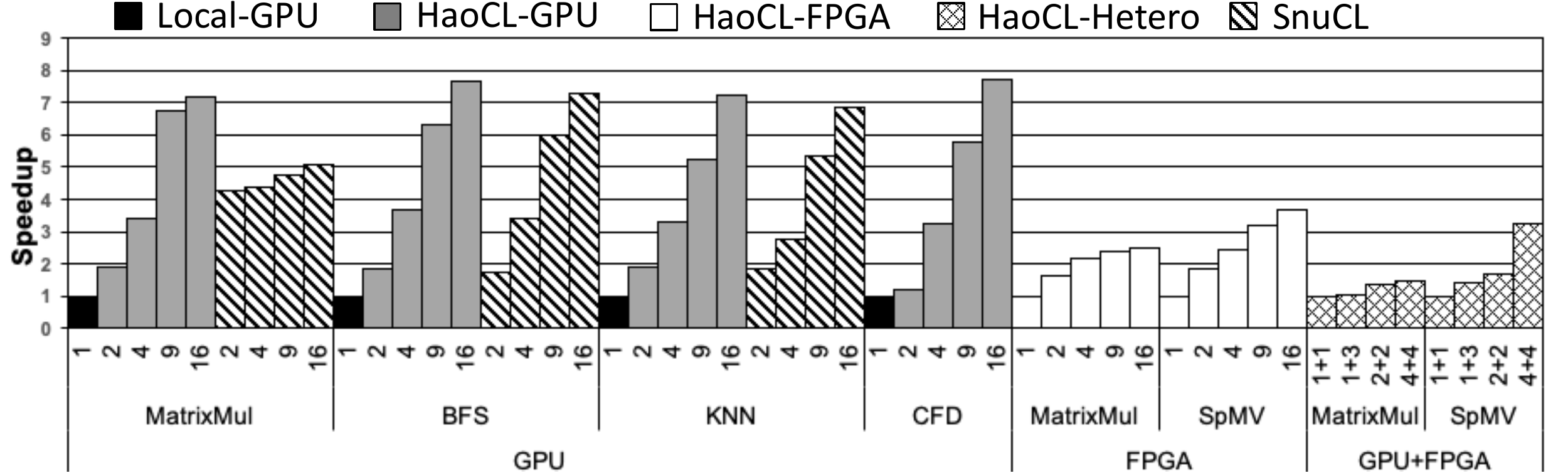}
    \vspace{-6mm}
    \caption{End-to-end speedup over a single GPU and FPGA.}
    \vspace{-2mm}
    \label{fig:e2e_speedup}
\end{figure}

MM and SpMv are chosen as benchmarks for the heterogeneity evaluation because of their data transmission complexity and the different requirement of efficiency. 
Their performance is normalized to a single node with FPGA or GPU, as shown in Fig.~\ref{fig:e2e_speedup}.
Noted that the MatrixMul kernels on the different devices are kept the same, just processing different data portion; while the different kernels (stages) of the SpMv are allocated to different devices, \textit{i.e.}, the kernel for data partition is allocated on the GPUs and computation on the FPGAs.
The system performance is linearly scaled with the number of the device nodes. 
Overall, the heterogeneity of the devices in the cluster is well utilized by leveraging the computation capacity of the nodes.

\subsection{Breakdown Analysis} \label{subsec:network}

\begin{figure}
    \centering
    \includegraphics[width=\columnwidth]{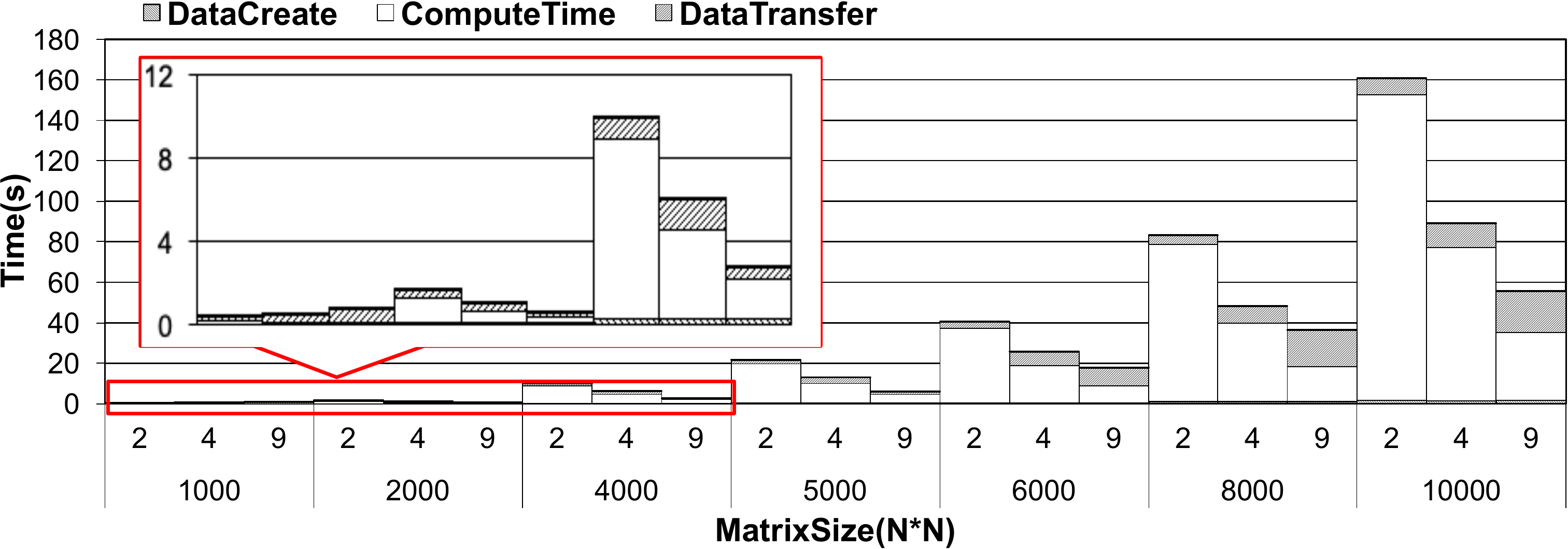}
    \vspace{-6mm}
    \caption{System breakdown analysis with Matrix Multiplication.}
    \vspace{-4mm}
    \label{fig:breakdown}
    
\end{figure}

Matrix Multiplication is conducted for a breakdown analysis of the HaoCL framework.
Since our node management process manages all the nodes with the same behavior, so the CPU, GPU and FPGA node have the same overhead of communication and local environment setup.
The breakdown analysis of a cluster with different number of GPU devices is shown in Fig.~\ref{fig:breakdown}.
The major runtime of the HaoCL framework includes system initialization time, data creation time, data transfer time and compute time.
Note that comparing to the later three, the system initialization time is negligible and is omitted in the Fig.~\ref{fig:breakdown}.
Although the communication and data creation time increase when the system and/or data size scale up, the ratio of them decreases.
\section*{Acknowledgement} \label{sec:ack}
This project is supported by Alibaba Innovation Research (AIR) grant. Besides, this project is also partly supported by the National Research Foundation, Prime Minister's Office, Singapore under its Campus for Research Excellence and Technological Enterprise (CREATE) programme.

\setstretch{0.75}
\newcommand{\BIBdecl}{\setlength{\itemsep}{0.25em}}
\bibliographystyle{IEEEtran}
\bibliography{IEEEabrv,Reference}
\end{document}